\documentclass[a4paper,twoside,preprint,english,aps,floatfix,superscriptaddress,showpacs]{revtex4}
\usepackage{times}
\usepackage[T1]{fontenc}
\usepackage[latin1]{inputenc}
\usepackage{amsmath}
\usepackage{graphicx}
\usepackage{amssymb}

\makeatletter

\usepackage{babel}
\makeatother

\begin{document}

\title{Spectral properties of a Rydberg atom immersed in a Bose-Einstein condensate}

\author{S. Middelkamp}

\email{stephan.middelkamp@pci.uni-heidelberg.de}

\affiliation{Theoretische Chemie, Institut f\"{u}r Physikalische Chemie, Universit\"{a}t
Heidelberg, INF 229, 69120 Heidelberg, Germany}

\author{I. Lesanovsky}

\email{Igor.Lesanovsky@uibk.ac.at}

\affiliation{Institut f\"ur Theoretische Physik, Universit\"{a}t Innsbruck, Technikerstraße 21 a, A-6020 Innsbruck, Austria}

\author{P. Schmelcher}

\email{peter.schmelcher@pci.uni-heidelberg.de}

\affiliation{Theoretische Chemie, Institut f\"{u}r Physikalische Chemie, Universit\"{a}t
Heidelberg, INF 229, 69120 Heidelberg, Germany}

\affiliation{Physikalisches Institut, Universit\"{a}t Heidelberg, Philosophenweg
12, 69120 Heidelberg, Germany}

\begin{abstract}
The electronic  spectrum of a Rydberg atom immersed in a Bose-Einstein condensate is investigated.
The Heisenberg equations of motions for the condensate and the Rydberg atom are derived. Neglecting the backaction of the Rydberg atom onto the condensate decouples the equations describing the condensate and  Rydberg atom. In this case the spectral structure of the Rydberg atom is completely determined by an effective potential which depends on the density distribution of the condensate. We study the spectral properties for the situation of an isotropic harmonic and anharmonic as well as axially symmetric confinement. In the latter case an intriguing analogy with Rydberg atoms in magnetic fields is encountered.
\end{abstract}

\date{\today}

\pacs{31.70.-f, 32.80.Rm, 03.75.Nt}

\maketitle
\section{Introduction}
The development of laser and evaporative cooling techniques in the last  decades has opened
up possibilities  to cool down alkali atoms to temperatures in the micro- to nanokelvin regime. In this regime the control over the external and internal degrees of freedom by external fields allow for the experimental realization and in the meantime routine production \cite{pethick,dalfovo99,pitaevskii} of Bose-Einstein condensates . Interacting Bose-Einstein condensates (BECs) exhibit nonlinear excitations such as solitons \cite{Denschlag2000a,Khaykovich2002a}, vortices \cite{Abo-Shaeer2001a,Fetter2001c} and are the origin of nonlinear matter-wave optics in general. Macroscopic as well as microscopic setups \cite{Folman2002,Fortagh2007a,Reichel2002} offer the possibility to coherently manipulate matter waves. 
Moreover Feshbach resonances \cite{Inouye1998a,marte02} allow for the control of the interaction properties. This enables one to create molecular BECs \cite{zwierlein03} and to probe  the BEC-BCS crossover \cite{Bartenstein2004} thereby opening up, together with quantum-phase-transition physics
in optical lattices, a strong link to condensed-matter physics.\\
Rydberg atoms, in contrast to atoms in their electronic ground states, possess extraordinary properties. They are of  gigantic extension and the highly excited states possess extremely long radiative lifetimes. Pictorially speaking the valence electron of a Rydberg atom is in a large loosely bound orbit. Consequently Rydberg atoms possess  large polarizabilities  \cite{Gallagher} and are extremely sensitive to external fields. Exposing them to a magnetic field they represent paradigm systems for classical and quantum chaos as well as modern semiclassics \cite{friedrich89}. A prominent example are the modulation peaks in photoabsorption spectra known as quasi-Landau resonances \cite{garton69} being  a signature of the dominant unstable periodic orbits in the classically chaotic regime.\\
There are different experimental techniques to trap long-lived cold Rydberg atoms being of optic \cite{Dutta2000}, magnetic \cite{choi05} or electric \cite{hyafil04} origin. Specifically for highly excited Rydberg atoms, their electronic structure cannot be neglected. Therefore it is crucial to study electronic properties of Rydberg atoms in inhomogeneous magnetic-field configurations putting aside the crude model of a point-like atom which is adequate for atoms in their electronic ground states. First investigations assumed a fixed nucleus taking advantage of the large atomic mass compared to the electronic mass. Major results observed  were interwoven spin-polarization patterns and magnetic-quadrupole-field-induced electric dipole moments induced by magnetic quadrupole fields \cite{lesanovsky04a,lesanovsky04b}. Taking into account the finite mass of the nucleus (ionic core) leads already in the presence of homogeneous magnetic fields to effects due to the non-separability of the center of mass and electronic motions \cite{schmelcher92,dippel94}. Trapping of Rydberg atoms in the quantum regime, i.e. a quantized center of mass motion coupled to the electronic state manifold, has been demonstrated very recently in case of the 3d quadrupole field for high angular momentum state \cite{lesanovsky05a} and  for the Ioffe-Pritchard configuration \cite{hezel06}.\\
Starting from an ultracold atomic cloud of ground state atoms, ultracold Rydberg gases can be created. Theses gases exhibit a  new kind of many-body physics \cite{killian99,pohl04}. Since the thermal motion can be neglected on the corresponding relevant short time scales, one can also speak of frozen Rydberg gases. Rydberg-Rydberg interaction, arising from the large polarizability and the large dipole moment of the Rydberg atoms, are of major importance for these systems. This interaction can be controlled by tuning external fields. In particular the strong dipole-dipole interaction  of the Rydberg atoms can lead to the so-called dipole blockade which inhibits the creation of a second excited atom in the vicinity of a Rydberg atom \cite{lukin01,tong04,singer04}. Due to these unique properties  ultracold Rydberg gases represent an intense topical field of research both theoretically \cite{pohl03} and experimentally \cite{killian01a}.\\
The problem we investigate in this paper bridges the gap between the situations described above, namely the cases of many 
Rydberg atoms and that of many ground state atoms:  We explore the spectral structure of a Rydberg atom immersed into a Bose-Einstein condensate, assuming that the dimension of the condensate is much larger than the size of the Rydberg atom. This system is within the above-outlined context of immediate physical interest and well-accessible within present-days ultracold experimental techniques. We start our investigations with the many body Hamiltonian. This Hamiltonian is expressed in  second quantization with the help of the field operators. We consequently derive coupled equations of motion for  atoms in the electronic ground state and in an excited state. Employing several approximations allows us to simplify and decouple these equations. Consequently the Gross-Pitaevskii equation appears for the ground state atoms. For the excited atom  the electronic energy levels depend on the global appearance and the symmetry of the density distribution of the ground state atoms. We investigate the cases of homogeneous, isotropic harmonic and axial symmetric harmonic density distributions of the ground state atoms figuring out a crucial dependence of the electronic spectrum on the shape of the density distributions.\\
In section II we start with the many-body Hamiltonian of the system and derive the Heisenberg equations of motion of the field operators for a Rydberg and a ground state atom, respectively. 
Section III contains our approximations and their justifications. We introduce contact potentials for the interactions and neglect the backaction of the Rydberg atom onto the condensate thus separating the corresponding equations of motion.
In Section IV the spectral properties of the Rydberg atom for different density distributions of the condensate are investigated. We study a homogeneous condensate, an isotropic harmonically and anharmonically confined condensate and a condensate with an axially symmetric density distribution.
Finally Section V provides the conclusions and a brief outlook.

\section{Hamiltonian and equations of motion}
In our model we consider atoms  consisting of  a point-like ion and a valence electron. Let us distinguish between ground state atoms with  non-excited valence electrons and Rydberg atoms with  highly excited valence electrons. Assuming that  only one atom is excited into a Rydberg state we separate the interaction of this highly excited atom with the ground state atoms into two parts: The excited electronic and the ionic collisional interaction. This decomposition is good for high-angular-momentum Rydberg quantum states. Furthermore we neglect mass correction terms due to the finite mass of the core so that the center-of-mass of the atom is located at the position of the ion.  
The many-body Hamiltonian is then given by
\begin{eqnarray}
H & = &\sum_{i} \frac{ \vec P^2_{i}}{2M} + \sum_{i} \frac{ \vec p^2_{i}}{2m_{e}} + \sum_{i} V^{e}(\vec r_{i}) + \sum_{i} V^{T}(\vec R_{i}, \vec r_{i}) + \frac{1}{2}\sum_{\genfrac{}{}{0pt}{1}{i,j}{i \neq j}} P^{0}_{ij} V^{AA}(\vec R_{i}-\vec R_{j}) P^{0}_{ij} \nonumber\\
&&  + \sum_{i,j}P^{1}_{ij}V^{IA}(\vec R_{i}- \vec R_{j}) P^{1}_{ij} + \sum_{i,j}P^{1}_{ij}V^{eA}(\vec R_{i},\vec r_{i},\vec R_{j}) P^{1}_{ij}  
\end{eqnarray}
where $M$, $m_{e}$ are the atom and the electron mass, respectively. $(\vec P_{i},\vec R_{i})$ denote the center of mass momentum and position vector for the $i$-th atom, respectively, while $(\vec p_{i},\vec r_{i})$ represent the corresponding momentum and coordinate vector for the valence electron with respect to the $i$-th parent ion. $V^{e}$ denotes the interaction of the ion with the valence electron which accounts in principle also for the effect of core scattering of the valence electron, e.g., the quantum defect. Later we will restrict ourselves to a pure Coulomb potential which correctly describes high angular momentum Rydberg quantum states.  The trapping potential $V^{T}$ is based on an adiabatic coupling of external fields to the magnetic moment or polarizability of the atom. $V^{AA}$, $V^{IA}$ and $V^{eA}$ are the interaction potentials between two ground state atoms (AA), an ion and a ground state atom (IA) and an excited electron and a ground state atom (eA), respectively. The projection operator $P^{0}$ ensures that the atom-atom interaction only occurs between two ground state atoms whereas $P^{1}$ ensures that there is only a contribution when a Rydberg atom interacts with a ground state atom.\\  
We define $\chi_{\alpha}(\vec r)$ to be the eigenfunctions of the electronic Hamiltonian $H^{e}=\frac{\vec p^2}{2m_{e}} + V^{e}(\vec r)$, i.e. the electronic states with the eigenenergies $E^{e}_{\alpha}$. For highly excited Rydberg atoms with the excited electron possessing a large angular momentum, these eigenfunctions are to a good approximation given by the hydrogen wave functions. We will focus here exclusively on this case and will label the hydrogen wave functions with the quantum numbers $(n,l,m)$ with their usual meanings.
The Hamiltonian in second quantization is then given by 
\begin{eqnarray}
H & = & \sum_{\alpha}\int d \vec R  \hat\Psi_{\alpha}^{\dagger}(\vec R) \bigr( -\frac{\Delta}{2M}+ E^{e}_{\alpha} \bigl) \hat\Psi_{\alpha}(\vec R) \nonumber\\
&&+ \sum_{\alpha,\beta}\int d \vec R \int d \vec r \hat\Psi_{\alpha}^{\dagger}(\vec R)\chi_{\alpha}(\vec r) V^{T}(\vec R, \vec r) \chi_{\beta}(\vec r) \hat\Psi_{\beta}(\vec R)\nonumber\\
&&+ \frac{1}{2} \int d \vec R  \int d \vec R^{\prime} \hat\Psi_{0}^{\dagger}(\vec R) \hat\Psi_{0}^{\dagger}(\vec R^{\prime}) V^{AA}(\vec R - \vec R^{\prime}) \hat\Psi_{0}(\vec R) \hat\Psi_{0}(\vec R^{\prime})\nonumber\\
&&+ \sum_{\alpha\not=0}\int d \vec R  \int d \vec R^{\prime} \hat\Psi^{\dagger}_{\alpha}(\vec R) \hat\Psi^{\dagger}_{0}(\vec R^{\prime}) V^{IA}(\vec R - \vec R^{\prime}) \hat\Psi_{\alpha}(\vec R)\hat\Psi_{0}(\vec R^{\prime}) \nonumber\\
&&+ \sum_{\alpha,\beta \not=0}\int d \vec R \int d \vec r \int d \vec R^{\prime}  \hat\Psi^{\dagger}_{\alpha}(\vec R) \hat\Psi^{\dagger}_{0}(\vec R^{\prime})\chi^{\star}_{\alpha}(\vec r) V^{eA}(\vec R,\vec r,\vec R^{\prime}) \chi_{\beta} (\vec r) \hat\Psi_{\beta}(\vec R)\hat\Psi_{0}(\vec R^{\prime})
\end{eqnarray}
where $\Psi_{\alpha}$ is the atomic field annihilation operator associated with an atom in the electronic state $\alpha$. Consequently $\Psi_{\alpha}(\vec R)$ describes an atom located at the position $\vec R$ in the electronic state $|\alpha\rangle$. In the above notation we suppressed the time argument.
One obtains via the Heisenberg equation  of motion for the ground state 
\begin{eqnarray}
i\partial_{t} \hat\Psi_{0}(\vec R) &=&-\frac{1}{2M} \Delta \hat\Psi_{0}(\vec R) + E^{e}_{0} \hat\Psi_{0}(\vec R) +  V^{T}(\vec R) \hat\Psi_{0}(\vec R)\nonumber\\
&&+  \int d \vec R^{\prime} V^{AA}(\vec R^{\prime} - \vec R) \hat\Psi_{0}^{\dagger}(\vec R^{\prime}) \hat\Psi_{0}(\vec R^{\prime})\hat\Psi_{0}(\vec R) \nonumber\\
&& + \sum_{\beta\not=0}\int d \vec R^{\prime} \hat\Psi_{\beta}^{\dagger}(\vec R^{\prime}) V^{IA}(\vec R^{\prime} - \vec R) \hat\Psi_{\beta}(\vec R^{\prime}) \hat\Psi_{0}(\vec R)\nonumber\\
&& + \sum_{\beta,\gamma \not=0}\int d \vec R^{\prime} \int d \vec r^{\prime}  \hat\Psi_{\beta}^{\dagger}(\vec R^{\prime})  \chi_{\beta}^{\star}(\vec r^{\prime}) V^{eA}(\vec R^{\prime},\vec r^{\prime},\vec R)\chi_{\gamma}(\vec r^{\prime}) \hat\Psi_{\gamma}(\vec R^{\prime})\hat\Psi_{0}(\vec R) \label{ground state exact}
\end{eqnarray}
and for an excited state
\begin{eqnarray}
i\partial_{t} \hat\Psi_{\alpha}(\vec R) &=& 
-\frac{1}{2M} \Delta \hat\Psi_{\alpha}(\vec R)+ E^{e}_{\alpha} \hat\Psi_{\alpha}(\vec R)\nonumber\\
&&+  \sum_{\beta} \int d r^{\prime} \chi_{\alpha}^{\star}(\vec r^{\prime}) V^{T}(\vec R,\vec r^{\prime}) \chi_{\beta}(\vec r^{\prime}) \hat\Psi_{\beta}(\vec R) \nonumber\\
&&+  \int d \vec R^{\prime} \hat\Psi_{0}^{\dagger}(\vec R^{\prime}) V^{IA}(\vec R - \vec R^{\prime}) \hat\Psi_{0}(\vec R^{\prime}) \hat\Psi_{\alpha}(\vec R)\nonumber\\
&&+ \sum_{\gamma \not=0}\int d \vec R^{\prime} \int d \vec r^{\prime}  \hat\Psi_{0}^{\dagger}(\vec R^{\prime})  \chi_{\alpha}^{\star}(\vec r^{\prime}) V^{eA}(\vec R,\vec r^{\prime},\vec R^{\prime})\chi_{\gamma}(\vec r^{\prime}) \hat\Psi_{0}(\vec R^{\prime})\hat\Psi_{\gamma}(\vec R) \label{excited state exact}
\end{eqnarray}

\section{Approximations: Contact Interaction and mean field approach}
Since we consider our gas to be ultracold, the velocities of the ground state atoms and of the ionic core of the Rydberg atom are very small, such that the potentials $V^{AA}$ and $V^{IA}$ are assumed to be determined by a single s-wave scattering length leading to the following parameterizations
\begin{eqnarray}
V^{AA}(\vec R - \vec R^{\prime}) &=& g\delta(\vec R - \vec R^{\prime})\\
V^{IA}(\vec R - \vec R^{\prime}) &=&\gamma\delta (\vec R  - \vec R^{\prime})
\end{eqnarray}
where $g$ and $\gamma$ are the corresponding energy-independent couplings. 
We assume the interaction between the excited electron and a ground state atom to be dominated by s-wave scattering. Being interested in principal effects we do neglect at this point  higher partial wave scattering which might be relevant in specific cases.  The velocity of the excited electron cannot be assumed to be  slow. Therefore we take as a potential for the interaction between the bound excited electron and a ground state atom a delta function with a scattering length depending on the kinetic energy of the electron. As an example we consider here $^{87}Rb$ according to ref. \cite{greene00}. 
\begin{figure}[htb]
\scalebox{1.0}{\includegraphics{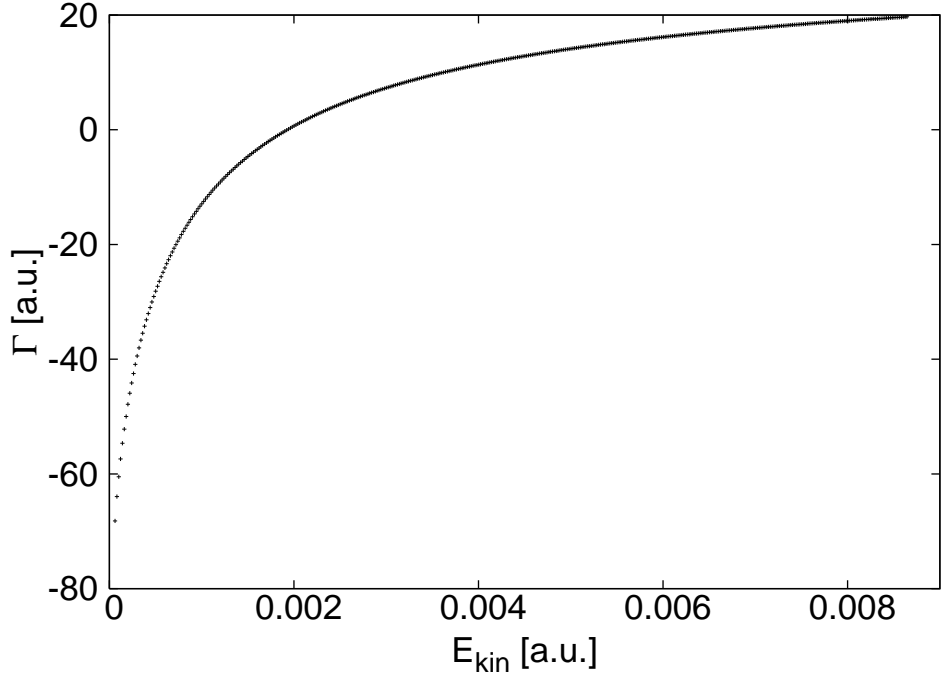}}
\caption{The dependency of the coupling  $\Gamma$ on the kinetic energy of the electron.}
\label{pahse shift}
\end{figure}
A relation between the distance of the electron from its parent ion and the velocity of the electron is provided by the classical relation $E_{kin}(r)=\frac{1}{2}k^2(r)=-\frac{1}{2n^2}+\frac{1}{r}$.
One can then determine the coupling  as a function of the distance $r$ by using the  triplet s-wave phase shift $\delta_{0}^{T}$ given in ref. \cite{hamilton02}.  
\begin{equation}
\Gamma(r)=-2\pi \frac{\tan(\delta_{0}^{T}(k))}{k}
\end{equation} 
This leads to the potential
\begin{equation}
V^{eA}(\vec R,\vec r,\vec R^{\prime}) = \Gamma(|\vec r|)\delta (\vec R + \vec r - \vec R^{\prime})
\end{equation}
For $r>2n^2$ where $E_{kin}$ would classically be negative we assume the coupling to be a constant, namely the corresponding value for a scattering electron in the zero kinetic energy limit. On the other hand side we assume that the Rydberg electron has an upper bound for its kinetic energy such that the available data (see Figure \ref{pahse shift}) on $\Gamma/\delta_{0}$ cover the complete collisional regime. The maximal kinetic energy is $E_{kin}^{max}=0.00864$ a.u. leading to a minimal distance $r_{min}=116$ a.u. The probability for an electron in a pure isolated Rydberg state, e.g., $n=30$ $l=15$ to be located in a distance from its parent ion smaller than $r_{min}$ is $0.0002$ and can therefore be neglected. Insertion of the potentials in eq. (\ref{ground state exact}) for the time evolution of the  ground state   leads to 
\begin{eqnarray}
i\partial_{t} \hat\Psi_{0}(\vec R) &=& 
\Bigl(-\frac{1}{2M} \Delta  + E^{e}_{0}  +  V^{T}(\vec R) +  g\hat\Psi_{0}^{\dagger}(\vec R) \hat\Psi_{0}(\vec R) 
 + \gamma\sum_{\beta\not=0} \hat\Psi_{\beta}^{\dagger}(\vec R) \hat\Psi_{\beta}(\vec R) \nonumber\\
&& + \sum_{\beta,\gamma \not=0}\int d \vec R^{\prime} \Gamma(|\vec R - \vec R^{\prime} |)  \hat\Psi_{\beta}^{\dagger}(\vec R^{\prime})  \chi_{\beta}^{\star}(\vec R-\vec R^{\prime}) \chi_{\gamma}(\vec R - \vec R^{\prime}) \hat\Psi_{\gamma}(\vec R^{\prime}) \Bigr)\hat\Psi_{0}(\vec R)
\end{eqnarray}
and in eq. (\ref{excited state exact}) for the excited state to
\begin{eqnarray}
i\partial_{t} \hat\Psi_{\alpha}(\vec R) &=&
\Bigl(-\frac{1}{2M} \Delta  + E^{e}_{\alpha}   
+  \gamma\hat\Psi_{0}^{\dagger}(\vec R) \hat\Psi_{0}(\vec R) \Bigr) \hat\Psi_{\alpha}(\vec R)\nonumber\\
&&+  \sum_{\beta} \int d r^{\prime} \chi_{\alpha}^{\star}(\vec r^{\prime}) V^{T}(\vec R,\vec r^{\prime}) \chi_{\beta}(\vec r^{\prime}) \hat\Psi_{\beta}(\vec R) \nonumber\\
&&+  \sum_{\gamma \not=0}\int d \vec R^{\prime} \Gamma(|\vec R^{\prime} - \vec R |) \hat\Psi_{0}^{\dagger}(\vec R^{\prime})  \chi_{\alpha}^{\star}(\vec R^{\prime}-\vec R) \chi_{\gamma}(\vec R^{\prime} - \vec R) \hat\Psi_{0}(\vec R^{\prime})\hat\Psi_{\gamma}(\vec R)
\end{eqnarray}
Since we restrict ourselves to the case of a single  Rydberg atom or several well isolated Rydberg atoms in the not-interacting approximation  one can substitute the field operator associated with an excited atom $\hat\Psi_{\alpha}(\vec R,t)$ by a wavefunction $\psi_{\alpha}(\vec R,t)$ where $\alpha$, as already indicated denotes the electronic state. It is important to note that the ansatz of a totally symmetric bosonic wave function of ground state atoms all residing in the same spatial orbital implies that any of these atoms can be excited in the course of the preparation of the system of a Rydberg atom plus condensate. Within our approach this property is reflected by the fact that we have an equally delocalized density of both the ground state atoms as well as the Rydberg atom (see further below).\\
Furthermore we assume the ground state to be occupied macroscopically such that we can introduce a mean field $\psi(\vec R,t)$ for the ground state atoms. This leads to
\begin{eqnarray}
i\partial_{t} \psi(\vec R,t) &=& 
\Bigl(-\frac{1}{2M} \Delta  + E^{e}_{0}  +  V^{T}(\vec R) +  g|\psi(\vec R,t)|^2  
+ \gamma\sum_{\beta\not=0}  |\psi_{\beta}(\vec R,t)|^2 \nonumber\\
&& + \sum_{\beta,\gamma \not=0}\int d \vec R^{\prime} \Gamma(|\vec R - \vec R^{\prime} |)  \psi_{\beta}^{\star}(\vec R^{\prime},t)  \chi_{\beta}^{\star}(\vec R-\vec R^{\prime})\nonumber\\
&&\chi_{\gamma}(\vec R - \vec R^{\prime}) \psi_{\gamma}(\vec R^{\prime},t) \Bigr)\psi(\vec R,t)\label{ground state}
\end{eqnarray}
and for the excited state to
\begin{eqnarray}
i\partial_{t} \psi_{\alpha}(\vec R,t) &=&
\Bigl(-\frac{1}{2M} \Delta  + E^{e}_{\alpha}   
+  \gamma |\psi(\vec R,t)|^2 \Bigr) \psi_{\alpha}(\vec R,t)\nonumber\\
&&+  \sum_{\beta} \int d r^{\prime} \chi_{\alpha}^{\star}(\vec r^{\prime}) V^{T}(\vec R,\vec r^{\prime}) \chi_{\beta}(\vec r^{\prime}) \psi_{\beta}(\vec R,t) \nonumber\\
&&+  \sum_{\gamma \not=0}\int d \vec R^{\prime} \Gamma(|\vec R^{\prime} - \vec R |)  \chi_{\alpha}^{\star}(\vec R^{\prime}-\vec R) \chi_{\gamma}(\vec R^{\prime} - \vec R) |\psi(\vec R^{\prime},t)|^2\psi_{\gamma}(\vec R,t) \label{excited state}
\end{eqnarray}
Eqs. (\ref{ground state},\ref{excited state}) represent coupled integro-differential equations for the mean field of the condensate and the center of mass amplitudes for the electronic states $\lbrace\alpha\rbrace$ of the Rydberg atom. Eq. (\ref{ground state}) contains the term of the traditional Gross-Pitaevskii equation augmented by terms that couple the mean field of the ground state atoms to the Rydberg atom. Eq. (\ref{excited state}) is the equation of motion for the center of mass amplitudes of the individual states $\lbrace\alpha\rbrace$ being coupled to the density of the mean field.\\
Eqs. (\ref{ground state},\ref{excited state}) represent a major challenge concerning their numerical solution. The simultaneous impact of both the 'background' coherent matter wave of ground state atoms and the trap potential, originating from electric, magnetic or electromagnetic fields on the Rydberg atom is of most delicate nature. Moreover since the focus of the present work is to reveal the impact of the matter wave background on the spectral properties of the electronic Rydberg wave function we will proceed in the following with a specification of the setup justifying the neglect of the terms involving $V^{T}$ in eq (\ref{excited state}). At first a condensate of ground state atoms is created. Then the trapping potential necessary for the formation of the condensate is switched off. Since the atoms in the condensate are ultracold one can neglect expansion effects of the atomic cloud on the timescale of the excitation of an atom into a Rydberg state via a laser. Due to this procedure one needs to consider the trapping potential in  eq. (\ref{ground state}) for the ground state atoms since the trap defines the shape of the condensate. However in eq. (\ref{excited state}) for the Rydberg atom the trapping potential should be absent. Because of the the slow expansion of the condensate cloud the situation after the excitation process can be regarded as a stationary one.   
The equation for the stationary solution with the chemical potential $\mu$ for the ground state atoms then reads 
\begin{eqnarray}
\mu \psi(\vec R) &=& 
\Bigl(-\frac{1}{2M} \Delta  + E^{e}_{0}  +  V^{T}(\vec R) +  g\psi(\vec R)^2
 + \gamma\sum_{\beta\not=0} \psi_{\beta}(\vec R)^2 \nonumber\\
&& + \sum_{\beta,\gamma \not=0}\int d \vec R^{\prime} \Gamma(|\vec R^{\prime} - \vec R|) \psi_{\beta}(\vec R^{\prime})  \chi_{\beta}^{\star}(\vec R-\vec R^{\prime}) \chi_{\gamma}(\vec R - \vec R^{\prime}) \psi_{\gamma}(\vec R^{\prime}) \Bigr)\psi(\vec R)
\end{eqnarray}
An estimation of the impact of the individual terms on the energy of the condensate with $N$ being the number of atoms leads to $ E_{AA}\sim g N^{2}$ for the energy resulting from the interaction of the ground state atoms with each other, $E_{IA}\sim \gamma N$ and $ E_{eA} \sim \Gamma N$ for the energy contributions resulting from the interactions of the ground state atoms with the ion and the excited electron, respectively. For a large number of ground state atoms in the condensate one can therefore neglect the contribution of $V^{IA}$ and $V^{eA}$ in comparison to the contribution of $V^{AA}$. The above approximations lead to a decoupling of the mean field equation for the condensate from the 'field' of the Rydberg atom. At the same time the equation for the Rydberg atom remains coupled to the condensate (see below, e.g., eq. (\ref{excited stationary})). Additionally we omit the energy offset due to the electronic energy of the ground state atoms. After introducing the  normalized  wavefunction $\phi_{0}(\vec R) = \frac{1}{\sqrt{N}}\psi(\vec R)$ and the density function $\rho(\vec R)=\psi(\vec R)^2 $ the usual Gross-Pitaevskii (GP) equation  appears for the ground state atoms \cite{pethick}.
\begin{eqnarray}
\mu \phi_{0}(\vec R) &=&\Bigl(-\frac{1}{2M} \Delta + V^{T}(\vec R) + g\rho(\vec R) 
\Bigr) \phi_{0}(\vec R)  \label{GP}  
\end{eqnarray}
Since we are in the present work interested in illuminating principal effects of ultracold clouds of ground state atoms on Rydberg atoms it is at this point appropriate to firstly employ the Thomas-Fermi approximation to the solution of the GP equation.
\begin{equation}
\rho(\vec R) = \frac{\mu-V^{T}(\vec R)}{g} \Theta(\vec R_{0}^2 - \vec R^2) \label{density TF} 
\end{equation}
The stationary equation belonging to eq. (\ref{excited state}) reads
\begin{eqnarray}
\epsilon \psi_{\alpha}(\vec R) &=& \sum_{\gamma \not=0} \int d \vec R^{\prime}  \chi_{\alpha}^{\star}(\vec R^{\prime}) \Bigl( -\frac{1}{2M} \Delta  + E^{e}_{\alpha}  
+  \gamma\rho(\vec R) \nonumber\\
&&+ \Gamma(|\vec R^{\prime}|)   \rho(\vec R^{\prime}+\vec R)\Bigr)\chi_{\gamma}(\vec R^{\prime})\psi_{\gamma}(\vec R)\label{excited stationary}
\end{eqnarray}
where the density $\rho$ is determined by eqs. (\ref{GP},\ref{density TF}). In order to proceed we make yet another crucial approximation. As discussed above we assume that the condensate wave function remains unchanged in the course of the switching-off of the trap and the excitation of the Rydberg atom. Now we additionally assume that the center-of-mass state of the Rydberg atom is the same as the one of the mean-field of the condensate. The motivation for this approximation is the fact that the original totally symmetric microscopic many boson wave function of ground state atoms leads to equal probability of Rydberg excitation for all atoms.
Consequently, the excitation process leads to a collective symmetric many-body quantum state  $|\psi_{e}\rangle=\frac{1}{\sqrt{N+1}}\sum_{i=1}^{N+1}|g\rangle_{1}|g\rangle_{2}|g\rangle_{3}\dots|g\rangle_{i-1}|e\rangle_{i}|g\rangle_{i+1}\dots|g\rangle_{N+1}$ where  $|g\rangle_{i}$ represents an atom numbered $i$ in the ground state and $|e\rangle_{i}$ an atom $i$ in a Rydberg state. It is therefore natural to assume that the center of mass amplitude for the Rydberg atom equals the corresponding normalized mean-field amplitude of the ground state atoms.\\ 
The above concept can be implemented by expanding the center-of-mass components $\psi_{\alpha}$ of the Rydberg atom in a basis set $\lbrace\phi_{i}(\vec R)\rbrace$ $\psi_{\alpha}(\vec R)=\sum_{i} \phi_{i}(\vec R) b_{i\alpha}$ with $\phi_{0}$ being the macroscopic wave function. Then all other expansion coefficients can be neglected compared to the expansion coefficient of $\phi_{0}$.
\begin{eqnarray}
\psi_{\alpha}(\vec R)=\sum_{i} \phi_{i}(\vec R) b_{i\alpha}\approx\phi_{0}(\vec R) b_{0\alpha} 
\end{eqnarray}
In the following we will suppress the index $0$ denoting the center-of-mass state.
Insertion of the above  approximation and projection on $\phi\equiv\phi_{0}$ leads to the matrix equation
\begin{eqnarray}
 \sum_{\gamma\not=0} H_{\alpha}^{\gamma} b_{\gamma}&=&\epsilon b_{\alpha}
\end{eqnarray}
with
\begin{eqnarray}
H_{\alpha}^{\gamma} &=& \int d \vec R \int d \vec R^{\prime} \phi(\vec R)\chi_{\alpha}^{\star}(\vec R^{\prime}) \Bigl( E^{e}_{\alpha} +  \gamma\rho(\vec R) +\Gamma(|\vec R^{\prime}|) \rho(\vec R^{\prime} +\vec R) \Bigr) \chi_{\gamma}(\vec R^{\prime}) \phi(\vec R) 
\end{eqnarray}
The first term in the above matrix is the electronic energy. Labelling the hydrogen wavefunctions $\chi_{\alpha}$ with the quantum number $(n,l,m)$ and assuming that $V^{e}(\vec r)$ is a pure Coulomb potential and/or focussing on higher angular momentum states in case of many-electron atoms with core scattering, $E_{\alpha}^{e}$ is a constant within a manifold of states belonging to the same principal quantum number $n$. We also focus on the case where the terms providing the mixing of the degenerate hydrogenic eigenstates do not induce mixing of states belonging to different $n$-manifolds. This implies that the interaction energy of the Rydberg atom with the ground state atoms is smaller than the spacing between these manifolds (see also section IV).  The second term results from the interaction of the ground state atoms with the ionic core. It does not depend on the Rydberg state and therefore cancels if one observes energy differences between two excited states. The last term results from the interaction of the excited electron with the ground state atoms and depends strongly on the electronic state. Consequently we will focus on this term. The corresponding matrix reads
\begin{eqnarray}
M_{\alpha}^{\gamma} &=& \int d \vec R \int d \vec R^{\prime} \phi(\vec R)\chi_{\alpha}^{\star}(\vec R^{\prime}) \Gamma(|\vec R^{\prime}|) \rho(\vec R^{\prime} +\vec R) \chi_{\gamma}(\vec R^{\prime}) \phi(\vec R) \label{matrix}
\end{eqnarray}
$M_{\alpha}^{\gamma}$ are the matrix elements of the effective potential
\begin{equation}
V(\vec R^{\prime})=\int d \vec R  \phi(\vec R)\Gamma(|\vec R^{\prime}|) \rho(\vec R^{\prime} +\vec R)  \phi(\vec R) \label{potential}
\end{equation}
in hydrogen wavefunctions. This potential can be varied in a wide range by changing the density distribution of the condensate due to the external trapping potential. As we shall exemplify in section IV the effective action of the condensate on the Rydberg atom via the potential (\ref{potential}) can lead, for example, to the case of a diamagnetic Rydberg atom if one chooses a strong two dimensional confining harmonic trapping potential


\section{Spectra of Rydberg atoms in a condensate}
\subsection{Homogeneous condensate}
We first consider as a trapping potential a rotational symmetric box of  radius $R_{0}$.
\begin{equation}
V^{T}(\vec R) = V_{0}(1-\Theta(\vec R_{0}^2-\vec R^2))
\end{equation}
This potential leads, for a large value of $V_{0}$, to a density distribution which increases from almost zero at the edge of the boxes to a constant on a scale of the healing length $\xi = \sqrt{\frac{1}{8\pi\rho_{0}a}}$ which is determined by the scattering length $a$ of the atoms. For a large enough box one can neglect these edge regions of non-constant density values and assume the density to be constant within the box and zero outside.
\begin{eqnarray}
\rho(\vec R)&=&\rho_{0}\Theta(\vec R_{0}^2-\vec R^2) \label{homogeneous density}
\end{eqnarray}
This leads with the number of particles being $N$ to the normalized wave-function $\phi(\vec R)=\sqrt{\frac{\rho}{N}}$.
Insertion of the density distribution (\ref{homogeneous density}) in eq. (\ref{matrix}) leads to
\begin{eqnarray}
M_{\alpha}^{\gamma} 
&=& \int d \vec R \int d \vec R^{\prime} \phi(\vec R)\chi_{\alpha}^{\star}(\vec R^{\prime}) \rho_{0}\Theta(\vec R_{0}^2-(\vec R+\vec R^{\prime})^2)\Gamma(|\vec R^{\prime}|) \chi_{\gamma}(\vec R^{\prime}) \phi(\vec R) \label{matrix homogen}
\end{eqnarray} 
The six-dimensional integrals (\ref{matrix homogen}) cannot be evaluated analytically and represent a major numerical challenge. However, a simple approximation render them tractable. Since the $\Theta$-function is a two-value piecewise constant function dropping the argument $\vec R^{\prime}$ in the $\Theta$-function just means that we assume the Rydberg atom alway to be completely 'covered' by the BEC. This leads to
\begin{eqnarray}
M_{\alpha}^{\gamma} 
&=& \int d \vec R \int d \vec R^{\prime} \phi(\vec R)\chi_{\alpha}^{\star}(\vec R^{\prime}) \rho_{0}\Theta(\vec R_{0}^2-\vec R^2)\Gamma(|\vec R^{\prime}|) \chi_{\gamma}(\vec R^{\prime}) \phi(\vec R) 
\end{eqnarray}
For an arbitrary large box this is obviously true but for a finite trap this is also a good approximation since we work in the limit where the typical dimensions of condensates are much larger than the size of the  Rydberg atoms. After inserting the center-of-mass wavefunction $\phi(\vec R)$  one can carry out the $\vec R$ integration by utilizing the normalization relation.
We label the hydrogen wave functions with the quantum numbers $n,l,m$. If one neglects $n$-mixing (see section III) the matrix is already diagonal since there are no angular dependent contributions.
\begin{eqnarray}
M_{n,l,m}^{n,l,m}&=&\frac{\mu}{g} \int dR^{\prime} R^{\prime2}R_{n,l}^2(R^{\prime})\Gamma(|\vec R^{\prime}|)\label{hom matrix}
\end{eqnarray}
The remaining radial integration was done numerically  by a Gaussian quadrature. 
\begin{figure}[htb]
\includegraphics{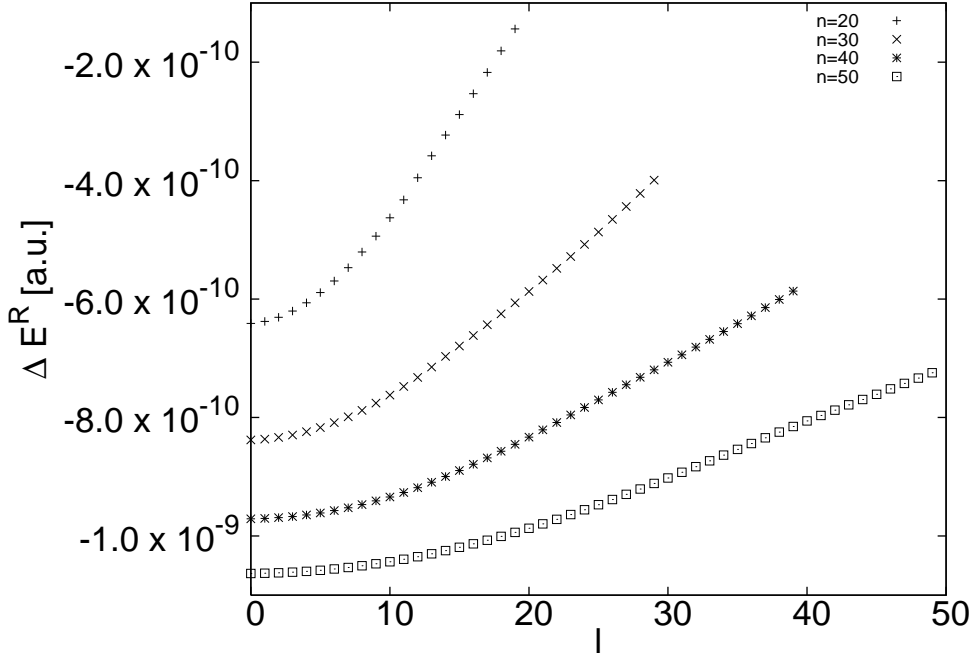}
\caption{Dependence of the energy shift $\Delta E^{R}$ in atomic units on the angular momentum $l$ for different $n$ for $\rho_{0}=1.5\cdot10^{-11}$ a.u.$\hat=10^{14}$ $\frac{1}{cm^3}$ for the case of a homogeneous condensate.}
\label{homogen} 
\end{figure}
Figure \ref{homogen} shows the dependence of the  shift $\Delta E^{R}$ of a certain state on its angular momentum $l$ for a typical density $\rho_{0}=1.5\cdot 10^{-11}$ a.u. This shift is negative for all states since the coupling constant of the electron-atom interaction is negative in the region where the spatial probability distribution of the electron has a maximum. The absolute value of the energy shift increases with increasing $n$ and decreases with increasing $l$. This can be explained by comparing the spatial probability distribution of the excited electron and the dependence of the coupling constant on the distance between the ion and the electron. The absolute value of the coupling $\Gamma$ increases with decreasing kinetic energy, which corresponds to a larger distance between the electron and the ion. At the same time, the expectation value of $R$ increases with decreasing $l$ for fixed $n$ and increases with increasing $n$ for fixed $l$. Therefore the absolute value of the energy shift decreases with increasing $l$. In the latter case one must additionally take into account that the spatial dependence of the coupling constant $\Gamma(-\frac{1}{2n^2}+\frac{1}{R})=-2\pi \frac{\tan(\delta_{0}^{T}(k))}{k}$ is different for different $n$. The energy shift does not depend on $m$, i.e. states with the same quantum numbers $n,l$ are degenerate.
For completeness we show in Figure \ref{homogen} all possible values for the angular momentum, although the strict validity of the contact interaction potential in s-wave form is only true for higher angular momenta which are also required if one wants to neglect core scattering.\\
Let us compare our predicted energy shift  with other typical energy scales of a Rydberg atom. 
The natural line widths due to spontaneous emission are typically  four orders of magnitude smaller for states with a large angular momentum than the typical predicted energy shifts. Line broadening due to the interaction of the electron with the ground state atoms can be neglected for ultracold atoms since the magnitude of the broadening depends on the induced phase shift which is small for ultracold atoms.
Relativistic effects such as spin-orbit coupling are about three orders of magnitude smaller. The pressure shift a Rydberg atom experiences in a cloud of non-condensed atoms is exactly the same as the one obtained above if one considers only s-wave scattering for the collisions between the non-condensed atoms and the Rydberg atom. 
\subsection{Isotropic harmonic and anharmonic confinement}
As a next trapping potential we consider an harmonic isotropic potential
\begin{equation}
V^{T}(\vec R) = \frac{1}{2}M\omega^{2} \vec R^{2}
\end{equation}
This leads to a density distribution  of the ground  state atoms according to eq. (\ref{density TF})
with $\vec R_{0}^{2}=\frac{2\mu}{M\omega^2}$ and a particle number of
\begin{eqnarray}
N&=&\frac{8\pi}{15}\bigl(\frac{2\mu}{M\omega^{2}}\bigr)^{\frac{3}{2}} \frac{\mu}{g} 
\end{eqnarray}
Insertion of the density distribution in eq. (\ref{matrix}) leads to
\begin{eqnarray}
M_{\alpha}^{\gamma} &=& \int d \vec R \int d \vec R^{\prime} \phi(\vec R)^2\chi_{\alpha}^{\star}(\vec R^{\prime}) 
\Gamma(|\vec R^{\prime}|) \Bigl(\frac{\mu-V^{T}(\vec R^{\prime} +\vec R)}{g}\Bigr) \Theta(\vec R_{0}^2 - (\vec R^{\prime }+\vec R)^{2}) \chi_{\gamma}(\vec R^{\prime}) \label{matrix with density}
\end{eqnarray}
Typical trap length scales are much larger than typical extensions of Rydberg atoms and the hydrogen radial wavefunctions vanish quickly for arguments larger than the dimension of the atom they describe.  Therefore we can simplify the argument of the $\Theta$-function $\Theta(\vec R_{0}^2 - (\vec R^{\prime }+\vec R)^{2})\rightarrow\Theta(\vec R_{0}^2 - \vec R^{2})$ since either $R^{\prime}\ll R_{0}$ or $R^{\prime}$ is larger than the extension of the Rydberg atom so that $\chi_{\alpha}(\vec R^{\prime})\approx0$. This simplification means again (see subsection IV.A) that the Rydberg atom is completely located within the condensate.
The center-of-mass function $\phi(\vec R)$ is defined as the normalized square root of the density.
Insertion of $\phi(\vec R)$ and of the hydrogen wave functions $\chi_{\alpha}(\vec R^{\prime})=R_{n,l}(R^{\prime})Y_{l,m}(\theta^{\prime},\phi^{\prime})$ makes it possible to fulfill all but one integration in spherical coordinates analytically. This leads for each $n$ manifold to the diagonal matrix
\begin{eqnarray}
M_{n,l,m}^{n,l,m} 
&=&  \frac{4\mu }{ 7  g }  \int_{0}^{\infty} dR^{\prime} R^{\prime2}  R_{n,l}^2(R^{\prime})\Gamma(R^{\prime})  \nonumber\\
&& - \frac{ M \omega^2}{2g}\int_{0}^{\infty} dR^{\prime} R^{\prime4}  R_{n,l}^2(R^{\prime})\Gamma(R^{\prime}) \label{harm matrix} 
\end{eqnarray}
This matrix is already diagonal because of the rotationally symmetric potential which leads to a conservation of the quantum numbers $l$,$m$. Again mixing of states with different quantum numbers $n$ need not  be taken into account because of their energy gap.\\
The energy shift $\Delta E^{R}$ of a state for $\omega\rightarrow0$ does, according to eq. (\ref{harm matrix}), not coincide with the energy shift of the homogeneous condensate in eq. (\ref{hom matrix}). This shows that the energy shift depends on the shape of the whole condensate and not only on the local density. The overall shape of a harmonically confined condensate is also for small trap frequencies $\omega$ different from the appearance of a homogeneous condensate. The dependence of the energy shift $\Delta E^{R}$ on the shape of the condensate results from the delocalization of the Rydberg atom within the condensate. 
\begin{figure}[htbp]
  \begin{minipage}[c]{8 cm}
\includegraphics[width=8cm]{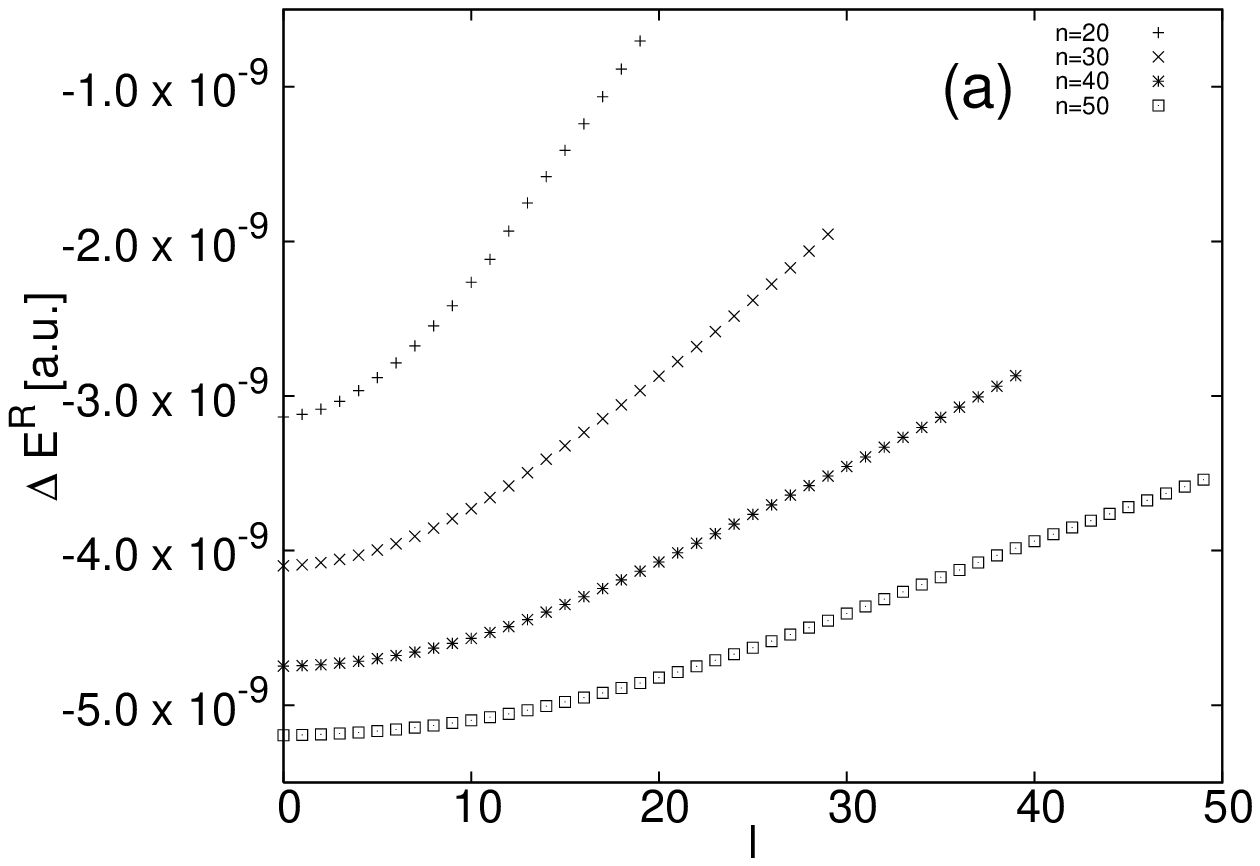}  
  \end{minipage}
  \begin{minipage}[c]{8 cm}
\includegraphics[width=8cm]{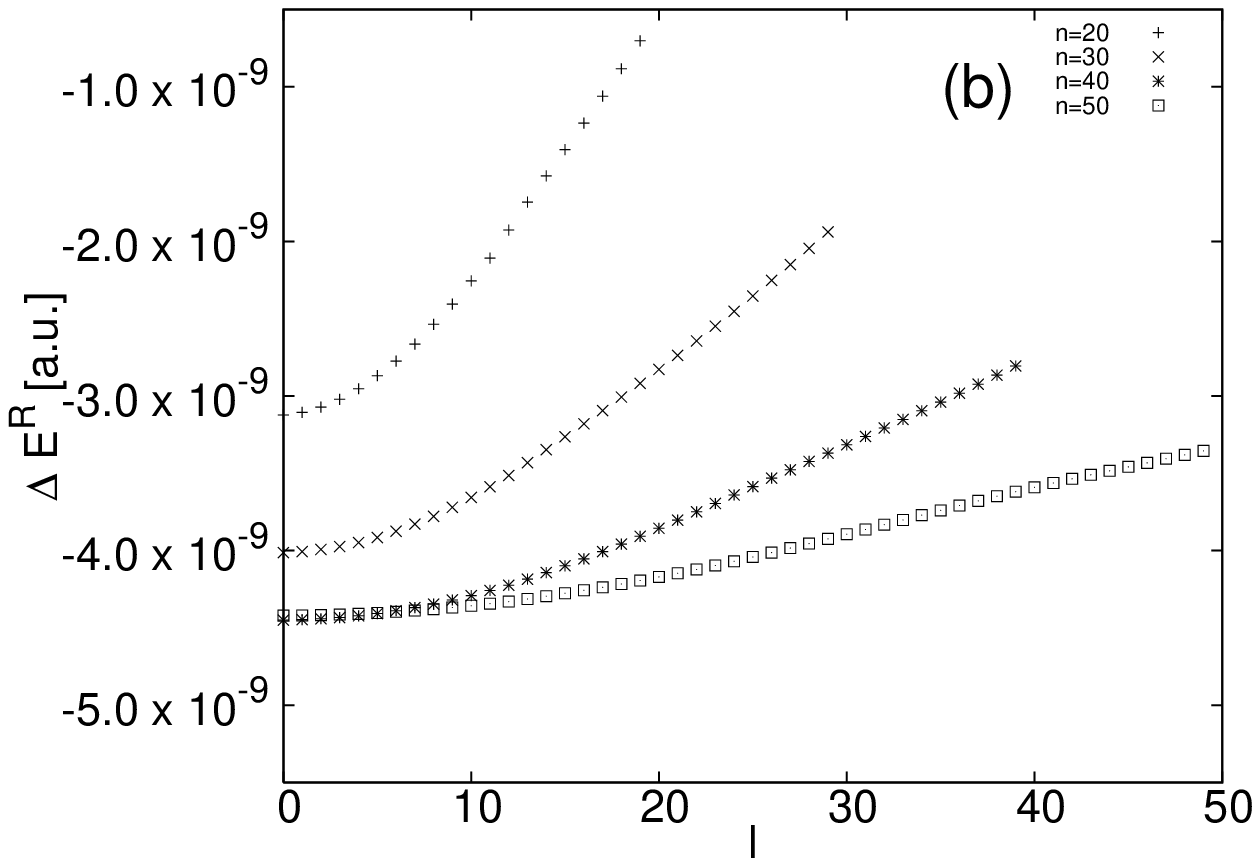}  
  \end{minipage}
  \caption{Dependence of the energy shift $\Delta E^{R}$ on the angular momentum $l$ for different $n$ for $\mu=10^{-12}$ for (a) $\omega =1$ kHz  and (b) $\omega = 10$ kHz.}
  \label{dep on l}
\end{figure}
Figure \ref{dep on l} shows the dependence of $\Delta E^{R}$ on the angular momentum $l$ for different $n$ for $\mu=10^{-12}$ and $\omega=1$ kHz or $\omega=10$ kHz. Out of convenience we denote the trapping frequency $\omega$ in Hz whereas $1$kHz$\hat=2.42\cdot10^{-17}$ a.u. For $\omega=1$ kHz the first term  in eq. (\ref{harm matrix}) dominates the energy shift. This summand is, apart from a factor $4$/$7$, equal to the energy shift of the homogeneous condensate. For $\omega = 10$ kHz the magnitude of the energy shift is, compared to $1$ kHz, smaller for sufficiently large quantum numbers $n$. For large quantum numbers $n$ the second summand in eq. (\ref{harm matrix}) becomes  important since the expectation value of $R^2$ increases according to the fourth power of $n$.
\begin{figure}[htb]
\scalebox{1.0}{\includegraphics{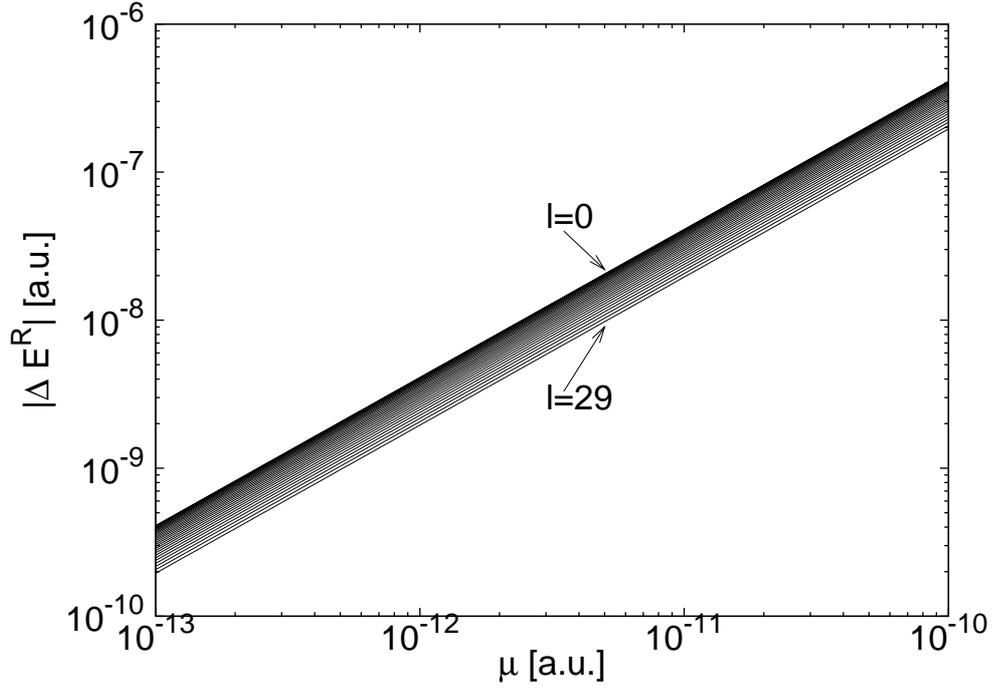} }
\caption{Absolute value of the energy shift $|\Delta E^{R}|$ of the Rydberg states with $n=30$ for an harmonically confined condensate with a trap frequency $\omega = 1$ kHz depending on the chemical potential $\mu$.}
\label{chem Pot omega = 1kHz}
\end{figure}
In Figure \ref{chem Pot omega = 1kHz} the dependence of the absolute value of the energy shift for the complete $n=30$ spectrum is shown as a function of the chemical potential $\mu$ for $\omega=1$ kHz. The absolute value of the energy shift increases for all states linearly  with the chemical potential. This  dependence results  from the linear dependence of the Thomas-Fermi density  distribution (\ref{density TF}) on the chemical potential. The slopes of the straight lines  that connect $|\Delta E_{R}|$ for states with the same angular momentum $l$ deviate only little.
\begin{figure}[htb]
\scalebox{1.0}{\includegraphics{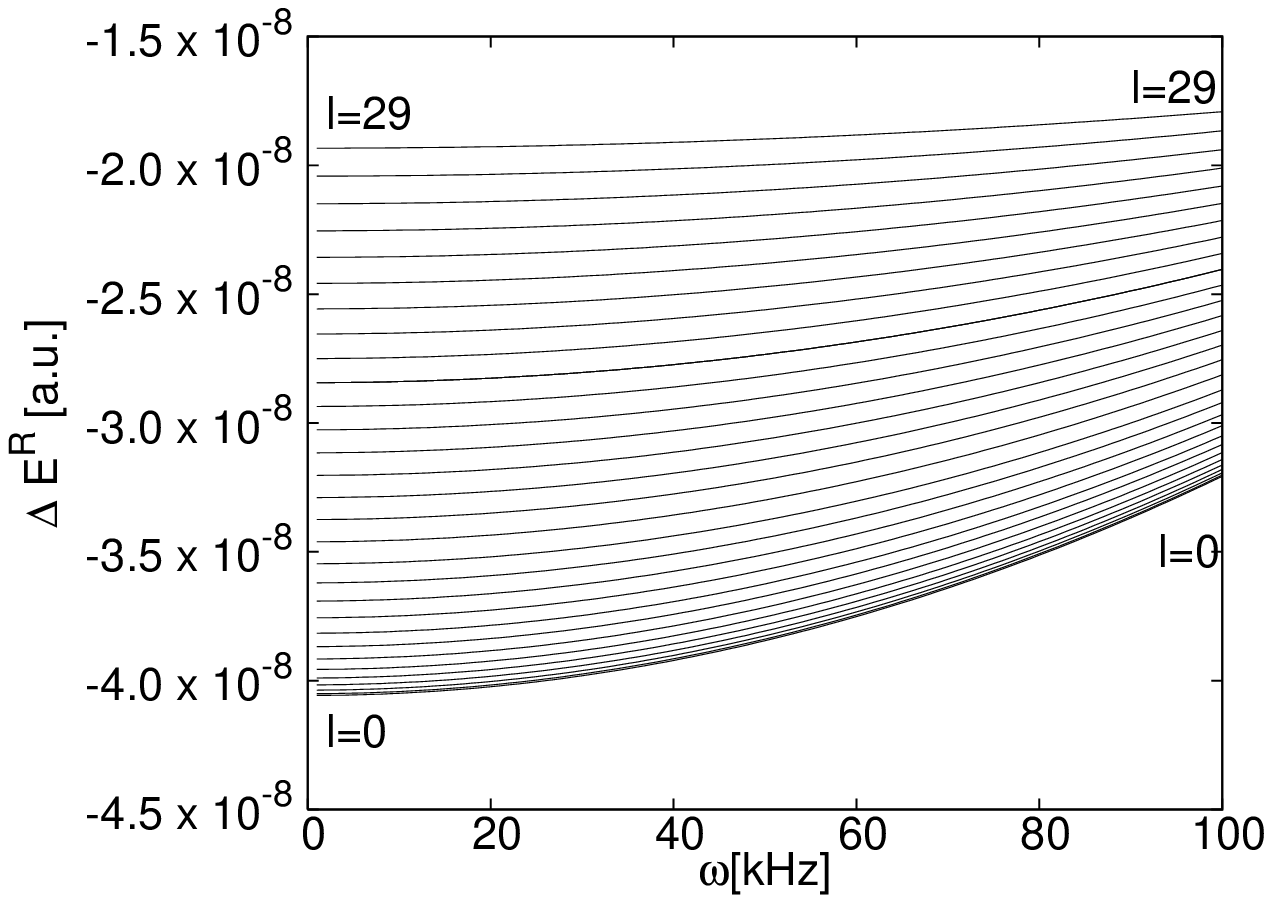}}
\caption{Energy shift of the Rydberg states with $n=30$ for an harmonically confined condensate with a chemical potential $\mu = 10^{-11}$ in dependence on the trap frequency $\omega$.}
\label{freq mu = 10^-11 }
\end{figure}
In Figure \ref{freq mu = 10^-11 } the dependence of the energy shift for the complete $n=30$ spectrum is shown as a function of the trapping frequency $\omega$ for $\mu=10^{-11}$. The absolute value of the energy shift decreases for all states quadratically  with the trapping frequency. The quadratic decrease can be explained by the quadratic decrease of the density with the trapping potential (see eqs. (\ref{matrix with density},\ref{harm matrix})).  The decrease  is different for the different states. States with  small angular momenta depend more strongly on the trap frequency than states with higher angular momenta.\\
For an anharmonic trapping potential $V^{T}(\vec R)=\alpha \vec R^4$ a similar argument as for the isotropic harmonic trap holds thereby leading to the matrix
\begin{eqnarray}
M_{n,l,m}^{n,l,m}
&=&\frac{8}{11}\frac{\mu}{g}\int_{0}^{\infty} dR^{\prime} R^{\prime2}  R_{n,l}(R^{\prime})\Gamma(R^{\prime})
R_{n,l}(R^{\prime})\nonumber\\
&&-\frac{7}{18}\frac{\sqrt{\mu\alpha}}{g}\int_{0}^{\infty} dR^{\prime} R^{\prime4}  R_{n,l}(R^{\prime})\Gamma(R^{\prime})
R_{n,l}(R^{\prime})\nonumber\\
&&-\frac{\alpha}{g}\int_{0}^{\infty} dR^{\prime} R^{\prime6}  R_{n,l}(R^{\prime})\Gamma(R^{\prime})
R_{n,l}(R^{\prime})
\end{eqnarray} 
Since the potential is rotationally symmetric the resulting matrix is diagonal, too. For a weak confining potential the energy shift is apart from a factor $8/11$ approximately the same as for the homogeneous case. For a stronger confining potential the second term is small compared to the third one since the expectation value of $R^6$ is for typical quantum numbers  $n,l$ of Rydberg atoms much larger than the corresponding expectation value of $R^4$. As a result one observes a linear decrease of the absolute value of the energy shift with increasing confining parameter $\alpha$ similar to the harmonic case.  
\subsection{Axially symmetric harmonic confinement}
Finally we investigate  an axially symmetric harmonic trapping potential 
\begin{eqnarray}
V^{T}(\vec R)&=&\frac{1}{2}M\omega^{2} (x^2+y^2)+\frac{1}{2}M\Omega^{2} z^2
\end{eqnarray}
The density profile is given by eq. (\ref{density TF})
with $R_{0}^{2}(\theta)=\frac{2\mu}{M\omega^2(1+ (\frac{\Omega^2}{\omega^2}-1) \cos^2(\theta))}$ whereas $\theta$ is the polar angle in spherical coordinates.
Spatial integration of the density leads to the number of particles
\begin{eqnarray}
N&=&\frac{8\pi\mu}{15g}\bigl(\frac{2\mu}{M\omega^2}\bigr)^{\frac{3}{2}} \frac{ \omega}{\Omega}   
\end{eqnarray}
This leads to the analogue of eq. (\ref{matrix with density}) with the corresponding density and the center-of-mass wave-function defined as its normalized square root. Here we can make a similar approximation to the isotropic confined condensate. We assume that the trap dimension in each direction is much larger than the size of the Rydberg atom and therefore the Rydberg atom is fully located within the condensate. Consequently the argument of the $\Theta$-function can be simplified and the integration of the center-of-mass coordinate can be accomplished analytically leading to 
\begin{eqnarray}
M_{\alpha}^{\gamma} &=&\frac{1}{g} \int_{0}^{\infty} dR^{\prime} R^{\prime2} \int_{-1}^{1} d\cos\theta^{\prime}  \int_{0}^{2\pi} d\phi^{\prime} \chi_{\alpha}^{\star}(R^{\prime},\theta^{\prime},\phi^{\prime})\Gamma(R^{\prime})\chi_{\gamma}(R^{\prime},\theta^{\prime},\phi^{\prime})\nonumber\\
&&\times\Bigl(\frac{4}{7}\mu- \frac{M \omega^2}{2}R^{\prime 2} - \frac{1}{2}M \cos^2\theta^{\prime} R^{\prime2} (\Omega^2-\omega^2) \Bigr) 
\end{eqnarray}
This matrix can be understood as an effective potential for the excited electron
 \begin{eqnarray}
V(\vec R^{\prime}) &=&\frac{\Gamma(R^{\prime})}{g} \Bigl(\frac{4}{7}\mu- \frac{M \omega^2}{2}R^{\prime 2} - \frac{1}{2}M \cos^2\theta^{\prime} R^{\prime2} (\Omega^2-\omega^2) \Bigr) \label{potential axial}
\end{eqnarray}
The following commutational and anticommutational relations hold with $P_{X^{\prime}}$ being the parity operator concerning $X^{\prime}$ and $L_{Z^{\prime}}$ the angular momentum operator concerning the $Z^{\prime}$ axis.
\begin{eqnarray}
\{P_{X^{\prime}},L_{Z^{\prime}}\}&=&0 \\
\lbrack V,L_{Z^{\prime}} \rbrack&=&0\\
\lbrack V,P_{X^{\prime}} \rbrack&=&0
\end{eqnarray}
If $|E,m\rangle$ is then an energy eigenstate and at the same time an eigenstate to $L_{Z^{\prime}}$ with $L_{Z^{\prime}}|E,m\rangle=m|E,m\rangle$
then we have
\begin{eqnarray}
L_{Z^{\prime}}P_{X^{\prime}}|E,m\rangle&=&-m|E,-m\rangle
\end{eqnarray}
Therefore $|E,m\rangle$ and $P_{X^{\prime}}|E,m\rangle=|E,-m\rangle$ are two degenerate energy eigenstates. So one expects for all states but states with $m=0$ a twofold degeneracy.
Doing the remaining angular integration with the help of the recursion relations of the Legendre Polynomials leads to the following non-zero matrix elements within a $n$ manifold
\begin{eqnarray}
M_{n,l,m}^{n,l,m}&=&\frac{4}{7}\frac{\mu}{g} \int_{0}^{\infty} dR^{\prime} R^{\prime2}  R_{n,l}^2(R^{\prime})\Gamma(R^{\prime})\nonumber\\
&&- \frac{M \omega^2}{2g}\Bigl(1+ \frac{(\Omega^2-\omega^2)}{\omega^2} \bigl(\frac{(l-m)(l+m)}{(2l+1)(2l-1)} +\frac{(l-m+1)(l+m+1)}{(2l+1)(2l+3)}\bigr)\Bigr)\nonumber\\
&& \times\int_{0}^{\infty} dR^{\prime} R^{\prime 4}  R_{n,l}(R^{\prime})\Gamma(R^{\prime})R_{n,l}(R^{\prime})  
\end{eqnarray}

\begin{eqnarray}
M_{n,l,m}^{n,l+2,m}&=&
- \frac{M(\Omega^2-\omega^2)}{2g} \sqrt{\frac{(l+m+2)(l+m+1)(l-m+1)(l-m+2)}{(2l+1)(2l+3)^2(2l+5)}}\nonumber\\
&& \times\int_{0}^{\infty} dR^{\prime} R^{\prime 4}  R_{n,l}^2(R^{\prime})\Gamma(R^{\prime})
\end{eqnarray}

\begin{eqnarray}
M_{n,l,m}^{n,l-2,m}&=&
- \frac{M(\Omega^2-\omega^2)}{2g} \sqrt{\frac{(l-m)(l-m-1)(l+m)(l+m-1)}{(2l+1)(2l-1)^2(2l-3)}}\nonumber\\
&& \times\int_{0}^{\infty} dR^{\prime} R^{\prime 4}  R_{n,l}^2(R^{\prime})\Gamma(R^{\prime})
\end{eqnarray}

For $\Omega=\omega$ the isotropic case is reproduced but in general the matrix is  no longer diagonal. Because of the axial symmetry $m$ remains a good quantum number. Only in the case of a near spherical symmetric trap the angular momentum $l$ represents an approximate constant of motion. In general their occurs a mixing of states with even and odd angular momentum $l$. As expected the  energy shift $\Delta E^{R}$  of a state depends on $|m|$, i.e. states where the absolute value of $m$ is different are no longer degenerate.  
We focus on states with $n=30$ and restrict our observations to a condensate with fixed chemical potential $\mu=10^{-11}$ since the dependence of the energy shift on the chemical potential is similar to the isotropic case.  
\begin{figure}[htb]
\scalebox{1.0}{\includegraphics{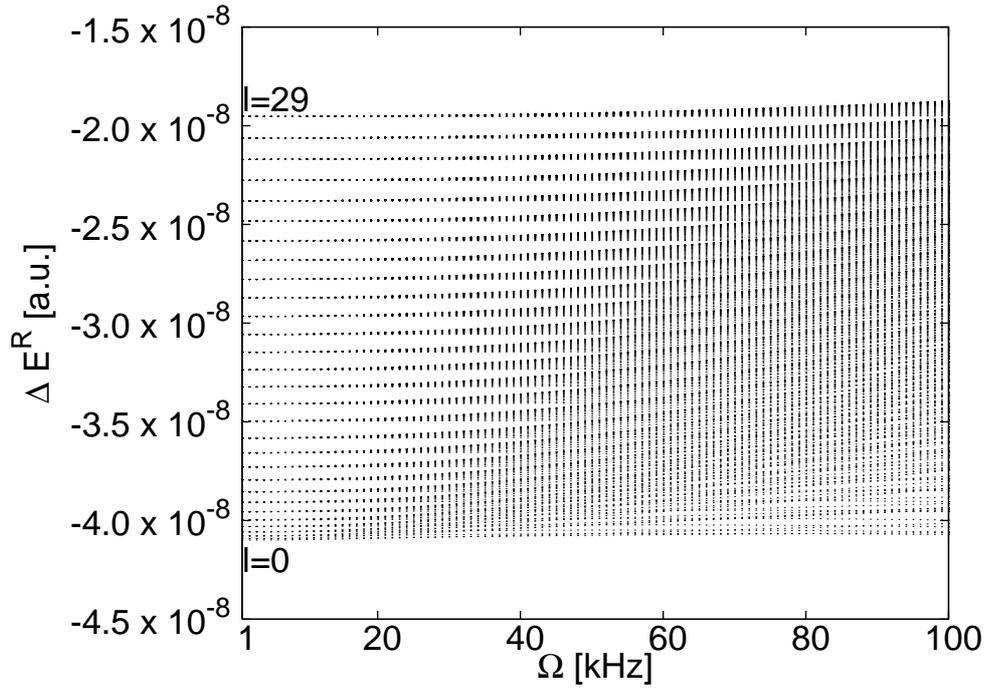}} 
\caption{Dependence of the energy shift $\Delta E^{R}$ for $n=30$ states  on the transversal trap frequency $\Omega$ for $\omega=1$ kHz and $\mu=10^{-11}$}
\label{all omega=1}
\end{figure}
In Figure \ref{all omega=1} the dependence of the energy shift on the transversal trap frequency $\Omega$ for $\omega=1$ kHz is shown. The degeneracy of the energy shift is reduced to a twofold one. The energy shifts for states with small angular momenta accumulate at small absolute values. Therefore $l$ is not a good quantum number. The energy shifts for high angular momentum states remain separated even for a large difference of the longitudinal and the transversal trap frequency. Therefore we will concentrate our investigations on states with $n=30$ and $l=29$.    

\begin{figure}[htbp]
  \begin{minipage}[c]{8 cm}
\includegraphics[width=8cm]{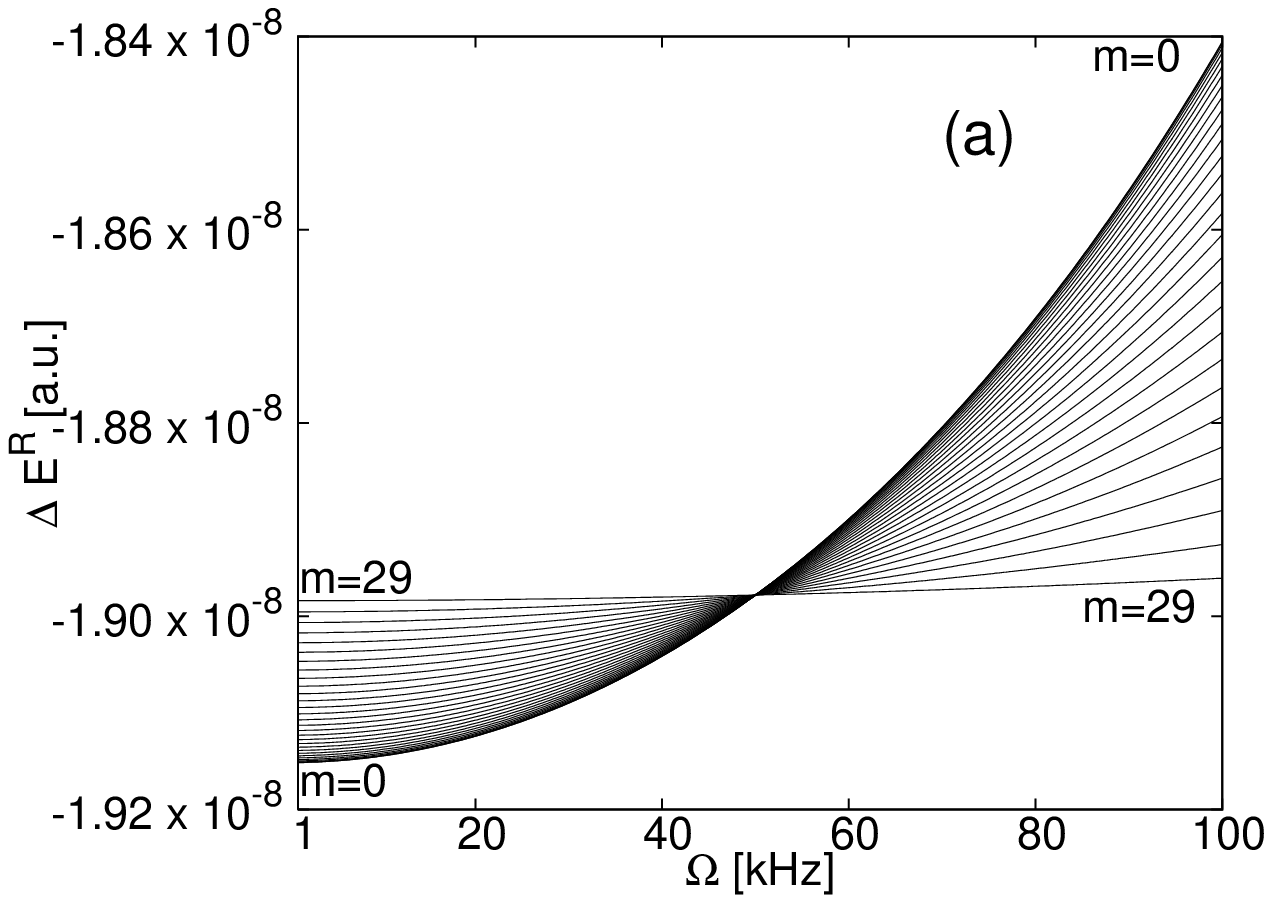} 
  \end{minipage}
  \begin{minipage}[c]{8 cm}
\includegraphics[width=8cm]{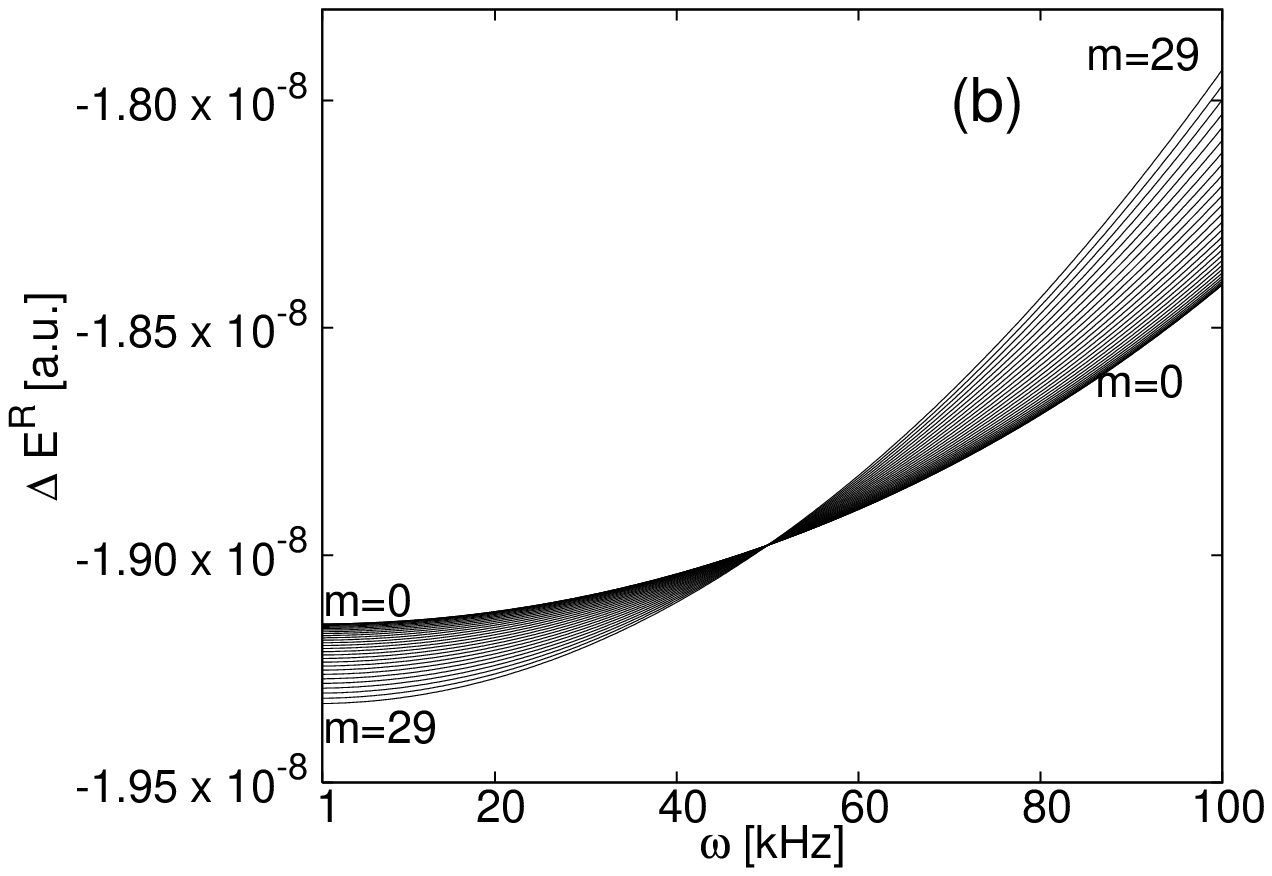}
  \end{minipage}
  \caption{Dependence of the energy shift of the $n=30$ $l=29$ states  on (a) the transversal trap frequency $\Omega$ for $\omega=50$ kHz and (b) the longitudinal trap frequency $\omega$ for $\Omega=50$ kHz with a chemical potential $\mu=10^{-11}$}
\label{l=29 omega=50} 
\end{figure}

Figure \ref{l=29 omega=50} (a) shows the dependence of the energy shift for the $n=30$ $l=29$ states on the transversal trap frequency.
The energy of the $m=29$ state is almost independent on the transversal trap frequency. This results from the fact that the states with extremal $m$ are located predominantly close to the $X$-$Y$-plane. Therefore the dependence of the energy shift on the transversal trap frequency decreases with increasing absolute value of the quantum number $m$ of the state.  Figure \ref{l=29 omega=50} (b) shows the dependence of the energy shift of the $n=30$ $l=29$ states on the longitudinal trap frequency. The absolute value of the energy shift decreases for all states  since $\omega$ defines the confinement in two dimensions, quadratically with $\omega$. Here the absolute value of the energy shift for the $m=0$ state decreases more slowly than the corresponding shift for larger $m$ since $m=0$ states are located around the $Z$-axis.

\begin{figure}[htb]
\scalebox{1.0}{\includegraphics{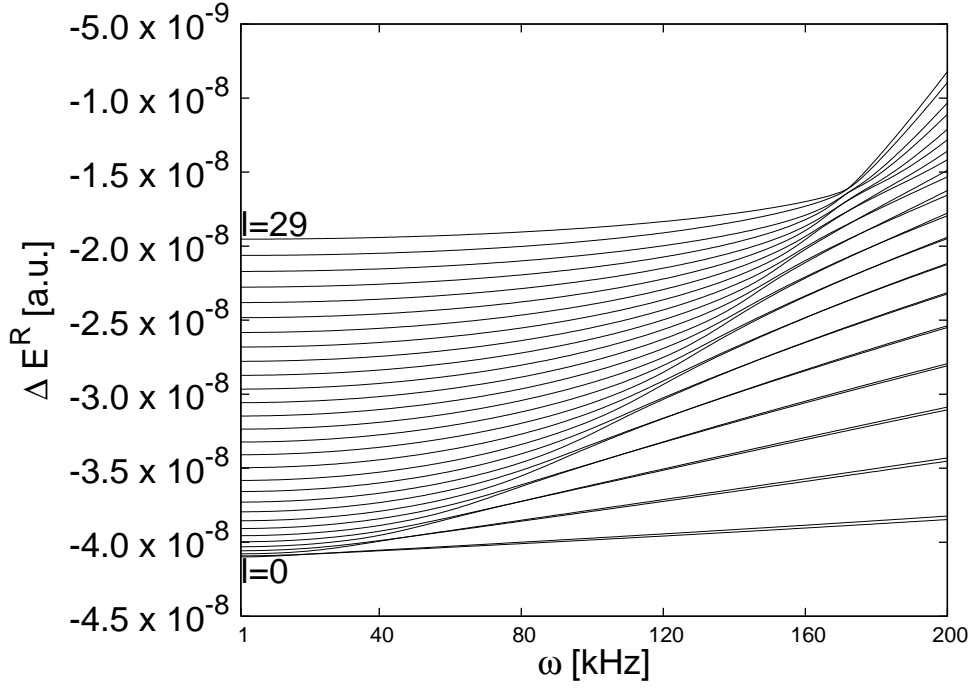}}
\caption{Dependence of the energy shift of the $n=30$ $m=0$ states on the longitudinal trap frequency $\omega$ for $\Omega=1$ kHz and $\mu=10^{-11}$}
\label{m=0 omega}
\end{figure}
Finally Figure \ref{m=0 omega} shows the dependence of the energy shift of the $m=0$ states. The $m=0$ states are for small trapping frequencies $\omega$ not degenerate since $l$ is no good quantum number.  For large trapping frequencies $\omega$ one recognizes in Figure \ref{m=0 omega} a rearrangement of the energy spectrum  leading to a pairing of the energy levels. Pairwise states with different parity which consist of a mixing of either exclusively even or exclusively odd angular momentum states experience almost the same energy shift. This can be explained by taking the limit of large trapping frequencies in the effective potential (\ref{potential axial}) leading to
\begin{equation}
V(\vec R^{\prime}) \sim \frac{ M \omega^{2} \Gamma( R^{\prime}) }{2g} R^{\prime 2} \sin^2\theta^{\prime}  
\end{equation}
The coupling coefficient $\Gamma(R^{\prime})$ is constant for large distances $R^{\prime}$ since the electronic energy is close to zero. The support of the square of the radial part of the hydrogen wave function times $R^{\prime 4}$  is for $n=30$ indeed in the regime where $\Gamma$ is constant as can be estimated by the expectation value of $R^{\prime 2}$ which is of order of $n^4$. In this limit $\Gamma=const.$ and for large $\omega$ the effective potential reduces to the diamagnetic interaction of an atom in a homogeneous magnetic field \cite{herrick82,friedrich89}. Correspondingly the resulting spectral structure observed in Figure \ref{m=0 omega} possess a completely analogous counterpart. The fact that a condensate Rydberg interaction can lead to effective potentials that are equally encountered for Rydberg atoms in external fields and/or mesoscopic environments is a intriguing analogy.

\section{Conclusions}
We have examined the electronic spectrum of a Rydberg atom immersed in a Bose-Einstein condensate of ground state atoms.
Starting with the many body Hamiltonian, we  expressed the Hamiltonian in second quantization and derive the Heisenberg equations of motions for the ground state atoms and an excited atom. For the mutual interaction of ground state atoms and  the interaction of the ionic core of the Rydberg atom and the ground state atoms we introduced  contact interactions with  constant couplings, respectively. The interaction of the Rydberg electron and the ground state atoms is modeled by s-wave scattering including the dependence of the coupling coefficient on the kinetic energy of the electron. The assumption of the macroscopic occupation of one state allowed us to introduce a mean field for the ground state atoms. Furthermore we  neglected the backaction of the Rydberg atom on the ground state atoms, thereby decoupling the equation of motion for the ground state atoms from the equation for the Rydberg atom.  Thus the Gross-Pitaevskii equation appeared for the ground state atoms. The equation for the Rydberg atom depends on the density distribution of the ground state atoms. A Rydberg atom in electric, magnetic or electromagnetic fields experiences a complicated trapping potential thereby coupling its center of mass and its electronic coordinates. In order to avoid the corresponding complications we assumed that the excitation of the Rydberg atom takes place after the trapping potential is switched-off. This is possible since the expansion velocity of a Bose-Einstein condensate without a confining potential is small so that the condensate can be regarded as static on a typical timescale it needs to excite an atom. We additionally assumed that the center-of-mass state of the Rydberg atom is the same as the one of the mean-field condensate. This is motivated by the fact that the original totally symmetric microscopic many boson wave function of ground state atoms leads to equal probability of Rydberg excitation for all atoms which suggests to use the same center of mass amplitude for the Rydberg atom as for the ground state atoms. This leads to a matrix which is analogue to the Hamiltonian matrix of a Rydberg atom in a potential. In our case the effective potential is provided by the density distribution of the condensate. We investigated the energy shift of a Rydberg state for different density distribution of the condensate. The energy shift  is always negative due to the coupling constant of the electron-atom interaction being predominantly negative in regions where the spatial probability distribution of the excited electron has a maximum. For  isotropic condensates the matrix describing the energy shift is already diagonal for each $n$-manifold since the quantum numbers $l$ and $m$ are conserved. Furthermore the energy shift does not depend on the quantum number $m$. The energy shift depends even for an homogeneous condensate on the quantum numbers $n$ and $l$ of the state of the Rydberg atom. Its absolute value increases with increasing quantum number $n$ and decreases with increasing quantum number $l$. For an isotropic harmonically confined potential we observed a linear increase of the absolute value of the energy shift with the chemical potential and a quadratic decrease with the trapping frequency of the condensate. For an axially symmetric shaped condensate the matrix is no longer diagonal and the degeneracy of the energy shifts is lifted apart from a twofold degeneracy for states with $m\neq0$. Furthermore the dependence of the energy shifts on the transversal and longitudinal trapping frequency is crucially different for different states. For a large transversal frequency the effective potential corresponds to the diamagnetic interaction of an atom in an homogeneous magnetic field. It is amazing that a condensate Rydberg interaction can lead to effective potentials analogous to potentials occurring for Rydberg atoms in external fields or more general to Rydberg atoms in mesoscopic environments.

\section{Acknowledgement}
Financial support by the Landesstiftung Baden-W\"urttemberg in the framework
of the project {}`Mesoscopics and atom optics of small ensembles
of ultracold atoms' is gratefully acknowledged by P.S. and S.M.
Fruitful discussions with J\"org Schmiedmayer are gratefully appreciated.

\bibliographystyle{prsty}
\bibliography{phd.bib,bec.bib}

\end{document}